\documentclass[reprint,
superscriptaddress,
 amsmath,amssymb,
 aps,
]{revtex4-1}

\usepackage{color}
\usepackage{graphicx}% Include figure files
\usepackage{dcolumn}% Align table columns on decimal point
\usepackage{bm}% bold math
\usepackage{graphicx}
\usepackage{epstopdf}
\usepackage{gensymb}
\usepackage{braket}
\usepackage{float}

\begin{document}

\title{Electronic correlations in the semiconducting half-Heusler compound FeVSb}

\author{Estiaque H. Shourov}
\affiliation{Materials Science and Engineering, University of Wisconsin--Madison}

\author{Patrick J. Strohbeen}
\affiliation{Materials Science and Engineering, University of Wisconsin--Madison}

\author{Dongxue Du}
\affiliation{Materials Science and Engineering, University of Wisconsin--Madison}

\author{Abhishek Sharan}
\affiliation{Department of Physics, University of Delaware}

\author{Felipe C. de Lima}
\affiliation{Department of Physics, University of Delaware}
\affiliation{Federal University of Uberlandia, Brazil}

\author{Fanny Rodolakis}
\affiliation{Advanced Photon Source, Argonne National Lab, Lemont, IL}

\author{Jessica McChesney}
\affiliation{Advanced Photon Source, Argonne National Lab, Lemont, IL}

\author{Vincent Yannello}
\affiliation{Department of Chemistry, University of Tampa, FL}

\author{Anderson Janotti}
\affiliation{Materials Science and Engineering, University of Delaware}

\author{Turan Birol}
\affiliation{Chemical Engineering and Materials Science, University of Minnesota Twin Cities}

\author{Jason K. Kawasaki}
\email{jkawasaki@wisc.edu}
\affiliation{Materials Science and Engineering, University of Wisconsin--Madison}

\date{\today}
\begin{abstract}

Electronic correlations are crucial to the low energy physics of metallic systems with localized $d$ and $f$ states; however, their effect on band insulators and semiconductors is typically negligible. Here, we measure the electronic structure of the half-Heusler compound FeVSb, a band insulator with filled shell configuration of 18 valence electrons per formula unit ($s^2 p^6 d^{10}$). Angle-resolved photoemission spectroscopy (ARPES) reveals a mass renormalization of $m^{*}/m_{bare}= 1.4$, where $m^{*}$ is the measured effective mass and $m_{bare}$ is the mass from density functional theory (DFT) calculations with no added on-site Coulomb repulsion. Our measurements are in quantitative agreement with dynamical mean field theory (DMFT) calculations, highlighting the many-body origin of the mass renormalization. This mass renormalization lies in dramatic contrast to other filled shell intermetallics, including the thermoelectric materials CoTiSb and NiTiSn; and has a similar origin to that in FeSi, where Hund's coupling induced fluctuations across the gap can explain a dynamical self-energy and correlations. Our work calls for a re-thinking of the role of correlations and Hund's coupling in intermetallic band insulators.

\end{abstract}

%\pacs{Valid PACS appear here}

\maketitle

\section{Introduction}

Electronic correlations are crucial for the low-energy properties of systems with highly localized $d$ and $f$ orbitals. Examples include correlated metals \cite{landau1980statistical}, Mott insulators \cite{mott1949basis}, Kondo systems \cite{hewson1997kondo}, and high temperature superconductors \cite{kamihara2008iron}. Although the effects of correlations are well established for metallic systems with partially filled bands, correlations in band insulators, for which there is a gap in the single particle spectrum, are generally overlooked due to the low carrier densities and the absence of low energy excitations. Correlated band insulators have been investigated theoretically using tight binding models \cite{garg2006can, kunevs2008temperature, sentef2009correlations}; however, beyond the exceptions of the narrow bandgap semiconductors ($E_g < 100$ meV) FeSi \cite{klein2008evidence, tomczak2012signatures} and Fe$_2$VAl \cite{nishino1997semiconductorlike, kristanovski2017quantum}, there are few well established real materials examples of correlated band insulators. Both Hubbard \cite{klein2008evidence} and Hund's \cite{tomczak2012signatures} couplings were shown to be capable of inducing correlations in band insulators, and the importance of Kondo description \cite{schlesinger1993unconventional} as well as spin fluctuations have been studied \cite{tomczak2012signatures, khmelevskyi2018correlated}.
%The microscopic mechanisms in these materials also remain under debate (Hubbard physics \cite{klein2008evidence}, Kondo \cite{schlesinger1993unconventional}, spin fluctuations \cite{tomczak2012signatures, khmelevskyi2018correlated}, and Hund's coupling \cite{tomczak2012signatures}) and the bandgaps are quite narrow ($\sim 50-100$ meV). 

Here we discover a new correlated band insulator: FeVSb. FeVSb crystallizes in the cubic half-Heusler structure and has a filled shell configuration of 18 valence electrons per formula unit. In a simple Zintl bonding picture, this corresponds to a filled [FeSb]$^{5-}$ polyanionic framework (Fe $d^{10}$, Sb $s^2 p^6$) with zincblende structure, and an empty cation V$^{5+}$ ($d^0$) that ``stuffs'' at the octahedral intersticials \cite{kandpal2006covalent} (Fig. 1). Density functional theory (DFT) calculations in the absence of on-site Coulomb repulsion predict a bandgap of 0.37 eV, larger than the $\sim 100$ meV predicted for FeSi \cite{arita2008, tomczak2013}. While FeVSb and other 18 electron half-Heuslers are promising materials for thermoelectric power conversion \cite{hinterleitner2019thermoelectric, zeier2016engineering, fu2015realizing}, Heusler compounds more broadly exhibit highly tunable topological states \cite{chadov2010tunable,lin2010half, manna2018heusler, hirschberger2016chiral}, magnetism \cite{wollmann2017heusler, palmstrom2003epitaxial, farshchi2013spin}, and novel superconductivity \cite{kim2018beyond, brydon2016pairing} as a function of electron count \cite{graf2011simple, kawasaki2019heusler}. Here, using angle-resolved photoemission spectroscopy measurements, we reveal a mass enhancement of $m^*/m_{bare} = 1.4$ in epitaxial FeVSb films with respect to the DFT band mass $m_{bare}$. This lies in striking contrast to other chemically similar $3d$ half-Heuslers, e.g., CoTiSb \cite{kawasaki2018simple} and NiTiSn \cite{ouardi2011symmetry}, for which photoemission and the bare DFT dispersions are in quantitative agreement. Our ARPES measurements for FeVSb are in quantitative agreement with realistic DFT + dynamical mean field theory (DMFT) calculations, suggesting that many-body correlations are essential to understanding its electronic structure. We compare with FeSi and comment on the possible role of Hund's coupling and spin fluctuations in enhancing the correlation strength.

\section{Results}

\begin{figure}[h]
    \centering
    \includegraphics[width=0.45\textwidth]{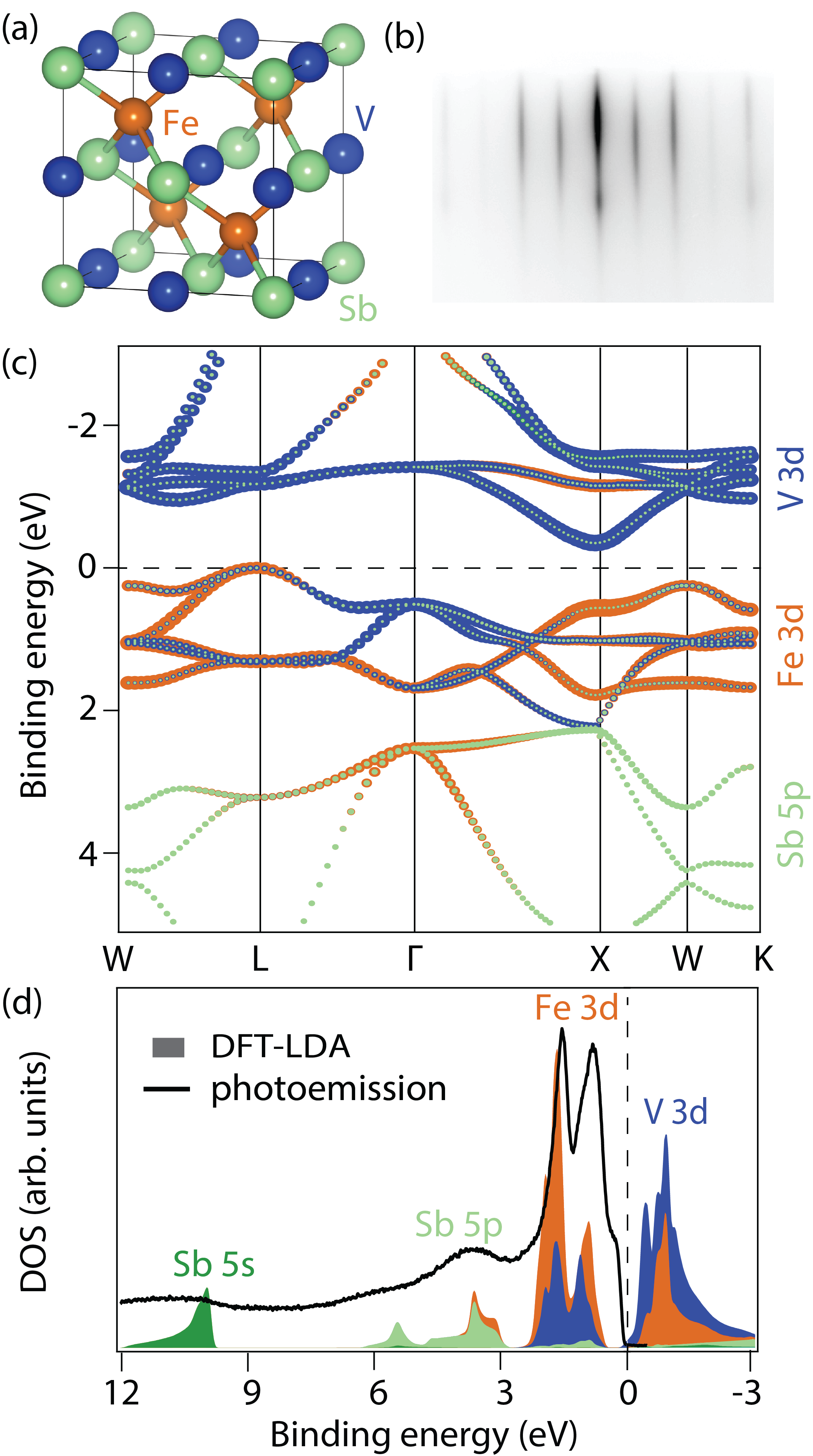}
    \caption{\textbf{Band character and atomic structure of FeVSb.} (a) Crystal structure. (b) Reflection high energy electron diffraction (RHEED) pattern. The sharp streaks indicate a smooth crystalline surface, suitable for photoemission measurements. (c) DFT-LDA calculated electronic structure of FeVSb and projected band character. (d) DFT-LDA ensity of states (shaded) and comparison to angle-integrated photoemission ($h\nu = 250$ eV). The calculated density of states has been shifted by 0.5 eV to match the measured valence band maximum from photoemission.}
    \label{dft}
\end{figure}

\textbf{General electronic structure of FeVSb.} As shown in Fig. \ref{dft}(c,d), the near $E_F$ bands of FeVSb have strong $3d$ character. In our DFT calculations using the local density approximation (LDA) with no added on-site electron-electron repulsion (+Hubbard $U$), the manifold of five valence bands just below the Fermi energy have primarily Fe $3d$ character and the 5-fold conduction bands have primarily V $3d$ character. The lower lying valence bands approximately 4 eV below the Fermi energy have Sb $5p$ character. The orbital character is nominally consistent with a Zintl bonding picture with the V $3d^0$ formally empty, and the Fe $3d^{10}$ and Sb $5s^2 5p^6$ formally filled \cite{kandpal2006covalent}. There is, however, significant Fe $3d$ - V $3d$ hybridization of the valence and conduction bands. Angle integrated measurements of the valence bands are consistent with this picture. Fig. \ref{dft}(d) shows an angle-integrated photoemission measurement of the valence band for our FeVSb film, grown by molecular beam epitaxy \cite{shourov2020semi} (Methods). At a 12 eV energy scale, the angle integrated spectrum is in qualitative agreement with the DFT-LDA calculations, with a one-to-one correspondence of the main Fe $3d$ and Sb $5p$ peaks expected.

\begin{figure}[h]
    \centering
    \includegraphics[width=0.45\textwidth]{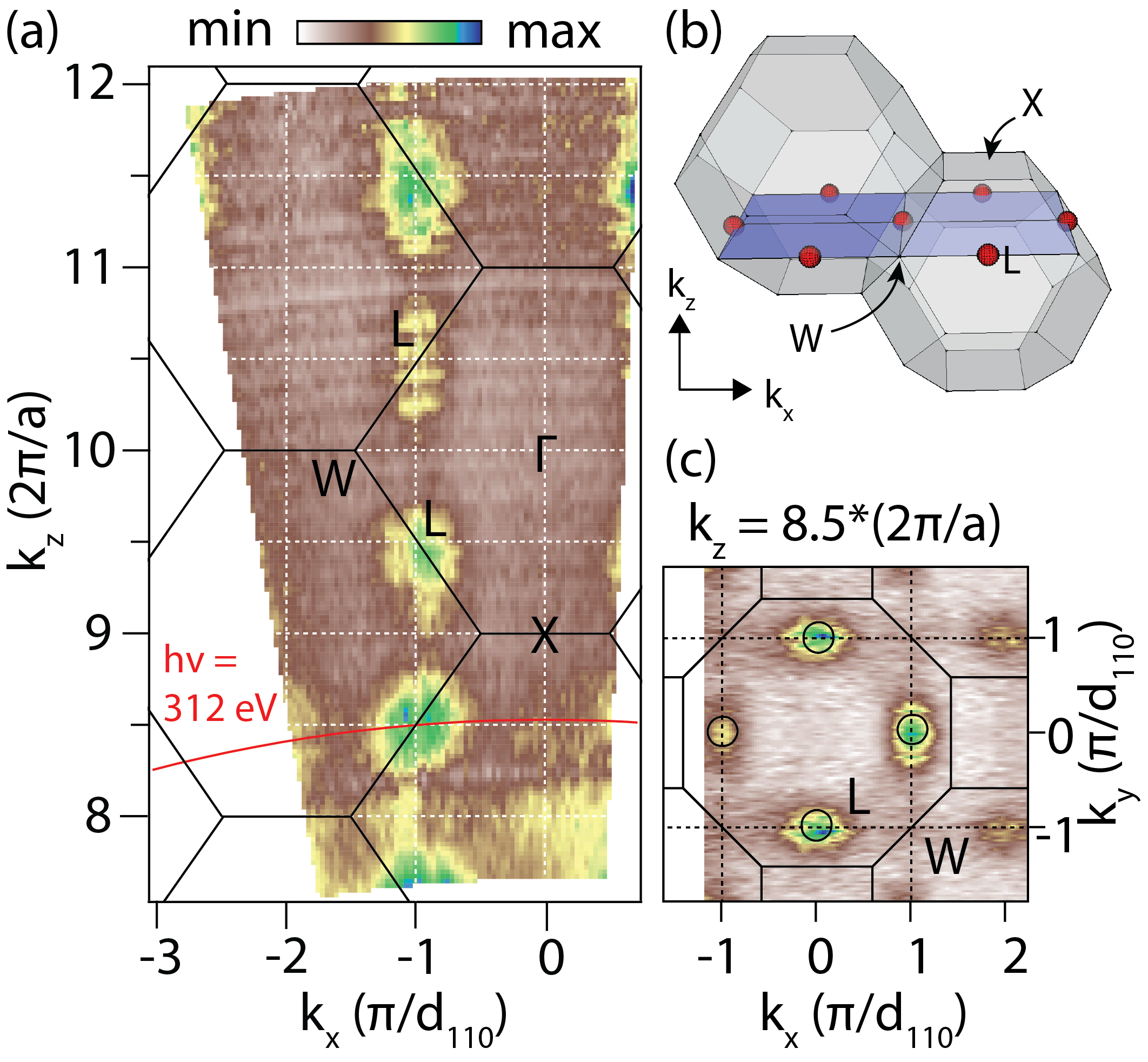}
    \caption{\textbf{Three-dimensional electronic structure of FeVSb.} (a) ARPES iso-energy cut through the top of the valence band ($E_B = 0.5$ eV, $k_y=0$), mapping the out-of-plane $k_z \parallel [001]$ dispersion. The color scale is the photoemission intensity. (b) Schematic iso-energy surface through the three-dimensional Brillouin zone. For clarity, only the $L$ centered hole pockets at $k_z = \frac{1}{2} (2\pi/a)$ are shown. Here, $k_x \parallel [110]$, $k_y \parallel [\bar{1}10]$, and $k_z \parallel [001]$. (c) In-plane constant energy slice at $h\nu=312$ eV, corresponding approximately to $k_z \approx  8.5 \times (2\pi/a) \rightarrow \frac{1}{2} (2\pi/a)$. The cut through the three-dimensional Brillouin zone is shown by the red arc in (a).}
    \label{kz}
\end{figure}

The general three-dimensional electronic structure is also in qualitative agreement with our DFT-LDA calculations. Fig. \ref{kz}(a) shows a measured iso-energy cut through the valence band maximum ($E_B = 0.5$ eV), tracking the out of plane $k_z$ dispersion. The data were compiled from photon energy dependent measurements from 250 to 695 eV, and $k_z$ was determined using a free-electron-like model of final states and an inner potential of $U_0=16$ eV to match the measured periodicity of bands. We observe hole pockets centered at the bulk $L$ points as expected from our DFT-LDA calculations. The in-plane $(k_x,k_y)$ iso-energy cut at a photon energy of 312 eV (Fig. \ref{kz}(c)), which corresponds approximately to a cut through constant $k_z \approx  8.5 \times (2\pi/a) \rightarrow  \frac{1}{2} (2\pi/a)$, is also in good agreement with DFT.

\begin{figure*}[t]
    \centering
    \includegraphics[width=1\textwidth]{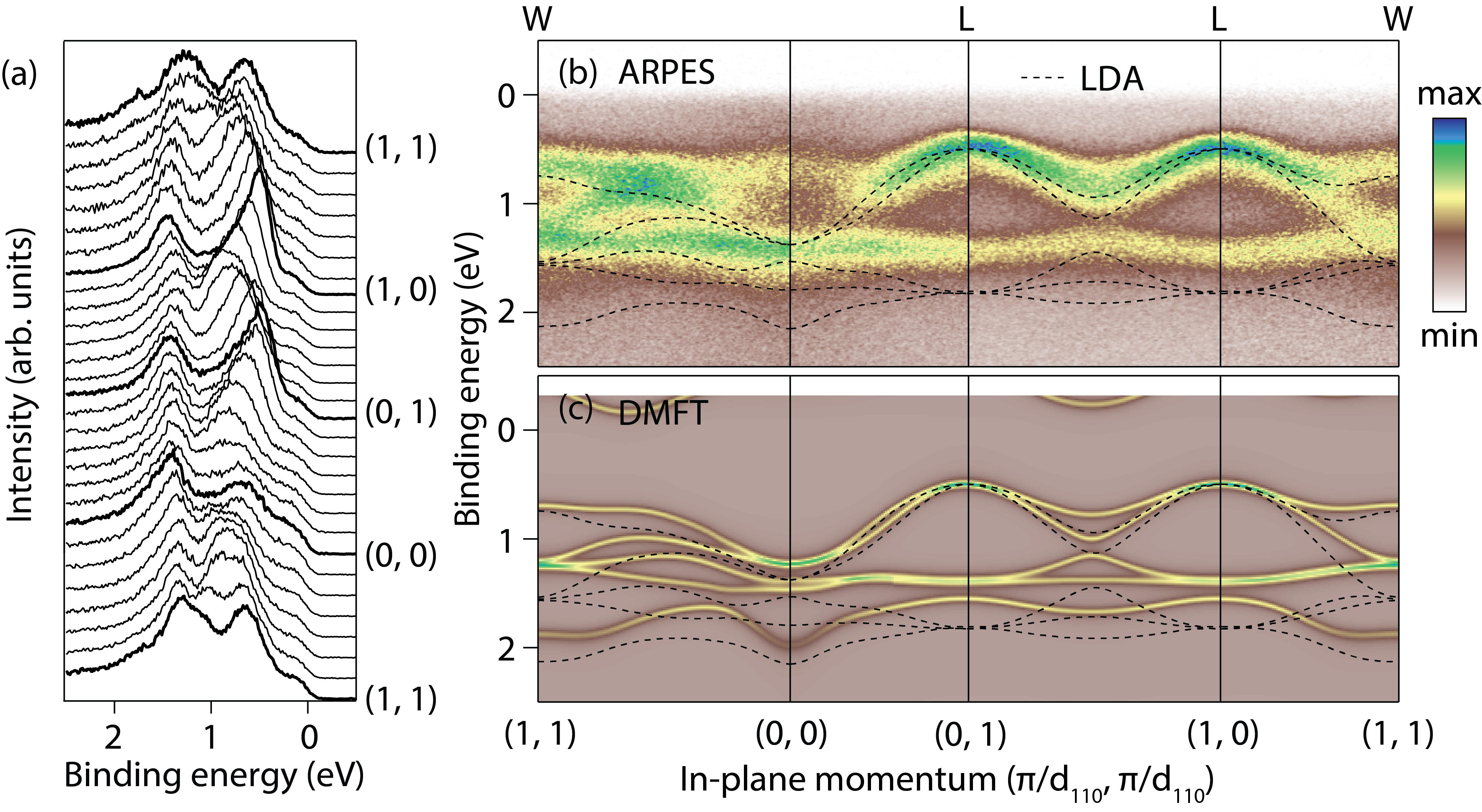}
    \caption{\textbf{Mass renormalization of FeVSb.} ARPES measurements in the $k_z \approx \frac{1}{2} (2\pi/a)$ plane ($h\nu=312$ eV), cutting through the bulk $L$ and $W$ points. (a) Energy dispersion curves (EDCs). (b) ARPES in-plane dispersion (color, same data as (a)), compared to a DFT-LDA calculation that does not include a Hubbard U (black curves). The measured dispersion is renormalized by a factor of $m^*/m_{bare} = 1.4$, where $m^*$ is the measured mass and $m_{bare}$ is the DFT-LDA mass. (c) Spectral function calculated by dynamical mean field theory (DMFT) with $U_{Fe}=U_{V}=4$ eV.}
    \label{arpes}
\end{figure*}

\textbf{Strong renormalization of the near-$E_F$ bands.} We now focus on in-plane energy dispersions, for which we observe significant band renormalizations. Fig. \ref{arpes} shows the dispersions through the high symmetry points in the $k_z \approx \frac{1}{2}(2\pi/a)$ plane, the same plane as shown in Fig. 2(b,c). The states at the bulk $L$ point [$(k_x, k_y)=(1,0)$ and $(0,1)$] form the top of the valence band. These states show an asymmetric lineshape in their energy dispersion curves (EDCs, Fig. \ref{arpes}(a)), which we attribute to $k_z$ broadening (Supplemental Fig. S-4). The measured ARPES dispersion (Fig. \ref{arpes}(b), color scale) is in qualitative agreement with the main features of the DFT-LDA calculation (black curves). However, we observe a narrowing of the measured electronic bandwidth $w$, or enhancement of the effective mass $m^*$, compared to the DFT bandwidth by a factor of $w_{bare}/w = m^*/m_{bare} = 1.4$. Here  $w_{bare}$ and $m_{bare}$ are the DFT bandwidth and effective mass, where we use the term ``bare'' since the DFT calculation was performed in the absence of on-site Coulomb repulsion (Hubbard $U=0$). 

\textbf{Ruling out extrinsic mechanisms for mass enhancement.} This mass enhancement for FeVSb cannot be explained by extrinsic mechanisms such as point defects, $k_z$ broadening, or strain (Supplemental). Briefly, ARPES measurements on an intentionally Fe-rich sample show that Fe antisite defects do not change the native dispersion. Rather, they simply add nondispersive spectral weight near $E_F$. $k_z$-broadening, due to the finite out-of-plane resolution of soft x-ray ARPES, does not significantly change the apparent dispersion based on simulated spectra with the experimental broadening $1/\lambda$, where $\lambda \approx 0.93$ nm is the photoelectron inelastic mean free path. Finally, the mass enhancement cannot be explained by strain, since our films are relaxed to the bulk lattice constant as measured by x-ray diffraction. Moreover, our DFT calculation show that in the absence of correlations an unphysically large strain of 10 percent would be required to produce the measured mass enhancement. 

\textbf{Electron correlations as the origin of mass enhancement.} The strong and intrinsic band renormalization suggests that an approach beyond DFT with local density or generalized gradient approximations (LDA or GGA) is needed to capture many-body exchange and correlation more accurately. A natural starting point is to consider hybrid functionals such as HSE \cite{heyd2003hybrid, krukau2006influence}, given the success of HSE in predicting the band gaps of compound semiconductors \cite{henderson2011accurate, garza2016predicting}. However, we find that HSE \textit{increases} the Fe $3d$ bandwidth at $L$ from 1.32 eV (GGA) to 1.5 eV (HSE), compared to the 0.94 eV bandwidth measured by ARPES (Supplemental Fig. S-1). DFT+U approaches also do not capture the measured reduction in bandwidth: values of $U$ on the Fe and V sites ranging from $U=0-4$ eV produced only moderate changes in the bandwidth, from 1.29 eV to 1.37 eV. These tests suggest that DFT approaches with static correlations cannot capture the low energy electronic structure of FeVSb.

For a more accurate treatment of many-body correlations, we turn to DFT plus dynamical mean field theory (DFT+DMFT). Unlike the hybrid functionals, DMFT reproduces \textit{dynamical} correlations that are local to an atomic site \cite{paul2019rev, kotliar2006, martin2016interacting}. We find quantitative agreement between ARPES and DMFT for $U_{Fe}=U_{V}=4$ eV (Fig. \ref{arpes}(c)). The value of $U$ is method and implementation dependent, and in our projector-based DMFT approach, where the $U$ is applied onto an atomic sphere, but not to a wider Wannier orbital, typically larger values of $U$ (as large as 10 eV for 3d transition metals) are used. The values of $U_{Fe}=U_{V}=4$ eV applied here are smaller than what has been shown to reproduce the the electronic structure of Fe pnictides or oxides \cite{haule2014covalency}, but similar in magnitude to the values that reproduces the electronic structure of FeAl alloys with other DMFT implementations ($U_{Fe}=3.36$ eV \cite{galler2015screened}). Tests with $U_{Fe}=6$ eV and $U_{V}=0$ produced similar renormalizations of the valence bands (Supplemental Fig. S-5). Our combined ARPES measurements and DMFT calculations suggest that dynamical correlations are essential for capturing the low energy electronic structure of FeVSb.

\section{Discussion}

Our observation of correlation-induced mass enhancement in FeVSb lies in striking contrast to chemically similar half-Heusler compounds such as CoTiSb \cite{kawasaki2018simple}, NiZrSn \cite{fu2019revealing}, and NiTiSn \cite{kawasaki2013epitaxial, ouardi2011symmetry}, for which photoemission and DFT calculations are in quantitative agreement ($m^*/m_{bare} = 1.0$). FeVSb, CoTiSb, and NiTiSn all have 18 valence electrons per formula unit, simple band theory predicts them to be diamagnetic semiconductors, and the valence bands all have strong $3d$ (Fe, Co, Ni) character. Additionally, our maximally localized Wannier function analysis \cite{abu2007ionicity} reveals that both FeVSb and CoTiSb share a similar spatial extent of the $3d$ orbitals and a similar degree of mixed covalent plus ionic bonding character (Supplemental Fig. S-6). Thus the fundamental question is: why is FeVSb correlated, while CoTiSb and NiTiSn are not?

We speculate that Hund's coupling may explain the enhanced correlations for FeVSb. Hund's coupling $J$ is known to strongly renormalize the electronic structure of ``Hund's metals,'' e.g., iron pnictides and ruthenates \cite{yin2011kinetic,georges2013strong, deng2019signatures}, and to a lesser extent the narrow bandgap semiconductor FeSi ($E_g \sim 100$ meV) \cite{tomczak2012signatures}. Our DMFT calculations reveal a qualitatively similar picture for FeVSb, despite its larger bandgap ($E_g \sim 0.4$ eV). For FeVSb we find the inverse of the quasiparticle residue $Z^{-1}$, calculated from slope of the frequency dependent electronic self energy $\Sigma(\omega)$ using \cite{han2016millis}
\begin{equation}
Z^{-1}=1-\frac{\partial \operatorname{Im}{\Sigma(i\omega_n)}}{\partial \omega}
\end{equation}
on the lowest Matsubara frequency $\omega_n$, is dependent on $J$. (In a Fermi liquid with linear self energy near $\omega=0$, $Z^{-1}$ is equal to the effective mass renormalization.) In particular, $Z^{-1}$ increases with $J$ in an orbitally selective manner, with $Z^{-1}=1.2-1.3$ at $J=0.7$ eV and $Z^{-1}=1.24-1.43$ at $J=1.0$ eV (Supplemental Fig. S-7). Neither this $J$-induced mass enhancement, nor the degree of its orbital selectivity, is as strong for FeVSb as it is for the prototypical Hund's metals such as Fe pnictides; but it is comparable to the semiconductor FeSi ($Z^{-1}=1.2-1.6$ at $J=0.7$ eV \cite{tomczak2012signatures}). 

In comparison, effects of Hund's coupling in CoTiSb are weaker than FeVSb, with $Z^{-1}=1.15-1.2$ at $J=0.7$ eV, and a weaker dependence on $J$. The weaker $J$ dependence in CoTiSb may be due, in part, to its larger bandgap (1.45 eV for CoTiSb, 0.37 eV for FeVSb), which makes the competition between the bandgap and the Hund's coupling go in the favor of the bandgap, suppressing correlation effects. Previous DFT calculations suggest that the bandgap for half-Heuslers follows a Zintl trend \cite{kandpal2006covalent}, in which the bandgap scales with the electronegativity difference between the two transition metals, e.g., Fe and V in FeVSb. From this trend the bandgap is expected to decrease across the series NiTiSn $\rightarrow$ CoTiSb $\rightarrow$ FeVSb, and the dependence of $Z^{-1}$ on $J$ is expected to increase.
%We caution that the strength of the $J$ dependence can depend on the magnitude of bandgap, which is difficult to predict from DFT or DFT+DMFT alone, since nonlocal static exchange has an effect on the bandgap as well, as shown by the HSE calculations.
We caution, however, that the bandgap is difficult to predict from DFT or DFT+DMFT alone, since nonlocal static exchange has an effect on the bandgap as well, as shown by the HSE calculations. More systematic studies are required to fully evaluate the effects of Hund's coupling and competition with the bandgap in these materials.

Spin fluctuations may also play a role in the enhanced correlations of FeVSb. Our DMFT calculations find that the expectation values for the \textit{magnitudes} of spin are larger for Fe and V in FeVSb ($\braket{|S_{Fe}|}= 0.73$, $\braket{|S_{V}|}= 0.65$) than Co and Ti in CoTiSb ($\braket{|S_{Co}|}= 0.62$, $\braket{|S_{Ti}|}= 0.54$), suggesting moderately larger spin fluctuations in FeVSb than CoTiSb. Since both compounds are band insulators in diamagnetic states in the absence of cross gap excitations, the marginally larger spin fluctuations may be an indication of the stronger correlations in FeVSb. 

In summary, we demonstrate electronic correlations are not limited to metals and narrow bandgap semiconductors. Our ARPES measurements reveal a mass enhancement of $m^*/m_{bare}=1.4$ in the semiconductor FeVSb, which is notable since FeVSb has a larger bandgap than previously identified correlated band insulators FeSi and Fe$_2$VAl. Hund's coupling may be responsible for the enhanced correlations in FeVSb, compared to chemically similar compounds. Generalizing the observations of correlations in FeSi to a system with a larger bandgap, our work shows that the Hund's coupling can affect dynamical correlation strength and bandwidth renormalization in semiconductors, as long as it is large enough to compete with the bandgap. Beyond the fundamentals implications on correlated electron systems, our discovery has a strong impact on applications such as thermoelectrics. For example, the thermoelectric power factor is highly sensitive to the effective masses \cite{pei2011convergence}, and spin fluctuations \cite{tsujii2019observation} are known to enhance the Seebeck coefficient. Correlated semiconductors with strong spin fluctuations are a promising platform for new thermoelectrics.

\section{Materials and Methods}

\textbf{MBE growth.} FeVSb films with thickness 50-100 nm were grown by molecular beam epitaxy on MgO (001) substrates. Samples were grown using a semi adsorption-controlled growth window, such that the Sb stoichiometry is self limiting. Further growth details are found in Ref. \cite{shourov2020semi}. Immediately following growth, samples were capped with $\sim 50$ nm Sb to protect the surface for transfer through air. Immediately prior to ARPES measurements, the Sb cap was desorbed in UHV by annealing the sample at 400$\degree$C, to reveal a clean FeVSb (001) surface. Surface cleanliness and full cap desorption were confirmed by photoemission core level measurements.

\textbf{ARPES measurements.} Angle resolved photoemission spectroscopy measurements were performed at beamline 29-ID of the Advanced Photon Source. We use horizontally polarized light with energy 250-1000 eV and a Scienta R4000 analyzer. All measurements were performed at a sample temperature of 20 K. The total energy resolution is $\sim 95$ meV and angular resolution 0.01 degrees. We determined the out-of-plane momentum using a free-electron-like model of final states $k_z = \sqrt{2m/\hbar^2} (E_{kin}cos^2 \theta + U_0)^{1/2}$. This model contains a single adjustable parameter, the inner potential $U_0$ which we determine to be $U_0=16$ eV by matching the periodicity of the photon energy dependent measurements.

\textbf{DFT calculations.} Density functional theory calculations were performed using the codes Wien2k and VASP. Calculations using the local density approximation (LDA), and the PBE0 \cite{perdew1996rationale} functional within the generalized gradient approximation (GGA) yield quantitatively similar bandstructures (Supplemental Fig. S-1). We also tested the Hybrid functional HSE06 \cite{krukau2006influence}. Further calculation details are found in the Supplement.

\textbf{Dynamical mean field theory (DMFT) calculations.} The DFT+DMFT calculations were performed using Rutgers eDMFT package \cite{Haule2010, Haule2016Forces} which uses the Wien2k code \cite{WIEN2K} for DFT, and the continuous time quantum Monte Carlo impurity solver \cite{Gull2011, Haule2007}. Exact double counting scheme \cite{Haule2015_double} was used, and the temperature was set to $\beta=1/50$~eV. In the calculations reported in the main text, both Fe and V atoms were treated as impurities, and all d orbitals were taken into account. (In some of the calculations reported in the supplemental information, only Fe was treated as an impurity.) Unless otherwise stated, $U=4$~eV and $J=0.7$~eV were used for both impurities. Note that due to the nature of the eDMFT implementation, the $U$ values used are typically higher than the Wannier based DMFT implementations. Time reversal symmetry was imposed in all DMFT calculations.

\bibliographystyle{apsrev}
\bibliography{bibliography}

\section{Acknowledgments}

We thank John Harter (UCSB) and Kyle Shen (Cornell) for the ARPES analysis tools. 

\textbf{Funding:} This work was supported by the CAREER program of the National Science Foundation (DMR-1752797) and the SEED program of the Wisconsin Materials Research Science and Engineering Center (DMR-1720415). This research used resources of the Advanced Photon Source, a U.S. Department of Energy (DOE) Office of Science User Facility operated for the DOE Office of Science by Argonne National Laboratory under Contract No. DE-AC02-06CH11357; additional support by National Science Foundation under Grant no. DMR-0703406. Work in the University of Minnesota was supported by the Department of Energy through the University of Minnesota Center for Quantum Materials under DE-SC-0016371. We acknowledge the Minnesota Supercomputing Institute for providing resources that contributed to the DMFT results reported within this paper. 

\textbf{Author contributions:} E.H.S. grew the samples. E.H.S., P.J.S., D.D., and J.K.K. performed the ARPES measurements with support from F.R. and J.M. E.H.S. and J.K.K. analyzed the ARPES data. A.S., F.C.L. and A.J. performed the DFT, HSE, and DFT+U calculations. V.Y. performed the Wannier function analysis. T.B. performed the DFT+DMFT calculations. E.H.S. and J.K.K. wrote the manuscript, with assistance from T.B. All authors edited the manuscript. J.K.K. conceived and supervised the project.

\end{document}